# Surface state evolution induced by magnetic order in axion insulator candidate EuIn$_2$As$_2$

Mingda Gong[1]*, Divyanshi Sar[1]*, Joel Friedman[1]*, Dariusz Kaczorowski[2], S. Abdel Razek[1], Wei-Cheng Lee[1]‡, and Pegor Aynajian[1]†

[1]Department of Physics, Applied Physics and Astronomy, Binghamton University, Binghamton, New York 13902, USA
[2]Institute of Low Temperature and Structure Research, Polish Academy of Sciences, 50-422 Wrocław, Poland

**Gapping of Dirac surface states through time reversal symmetry breaking may realize the axion insulator state in condensed matter. Despite tremendous efforts, only a few material systems fall in this category of intrinsic magnetic topological insulators (TI). Recent theoretical calculations proposed the antiferromagnetic EuIn$_2$As$_2$ to be a topologically non-trivial magnetic insulator with gapped surface states. Here we use scanning tunneling microscopy and spectroscopy (STM/STS) complemented with density-functional theory (DFT) calculations and modelling to probe the surface electronic states in EuIn$_2$As$_2$. We find a spin-orbit induced bulk gap of ~120 meV located only a few meV above the Fermi energy, within which topological surface states reside. Temperature dependent measurements provide evidence of the partial gapping (~40 meV) of the surface states at low temperatures below the AFM order, which decreases with increasing temperature but remains finite above T$_N$.**

Merging of magnetism with non-trivial topology induces some of the most fascinating phenomena in quantum materials from both a fundamental and practical perspective [1–8]. Examples include the quantum anomalous Hall effect, which manifests quantized Hall conductivity without an externally applied magnetic field [9–12], the topological axion insulator state with quantized magneto-electric effect [13,14], and Majorana fermions that obey non-Abelian statistics [15]. The magnetic phase transition in magnetic topological insulators (TI) offers the possibility to switch on/off the time reversal symmetry, providing tuning capabilities of their topological properties [6]. Vigorous research focus has thus been devoted to the interplay of magnetism and non-trivial topology [6,16–29], which due to dissipationless transport has transformative potential in quantum information technologies, overcoming power losses induced by the strong electron scatterings in trivial systems [30,31].

Despite the efforts, stoichiometric magnetic topological quantum materials remain rare [32–36]. One example that has recently attracted particular attention is the antiferromagnetic $MnBi_2Te_4$. Theoretical and experimental studied indicate topologically non-trivial surface states induced by its large spin orbit coupling [10,19,37–39]. Angle resolved photoemission spectroscopy (ARPES) experiments reveal a large gap of ~100 meV within the surface states in the antiferromagnetic (AFM) phase, believed to be related to the magnetic transition [40]. Yet the gap, which is located around ~ -300 meV, well below the Fermi energy, remains finite above $T_N$ making its association to the time reversal symmetry breaking dubious [40–42]. A possible interpretation of the observed insensitivity to magnetism is the weak hybridization of the magnetic Mn d-states with the Te p-states [10].

Very recently, $EuIn_2As_2$ has been theoretically proposed to host exotic topological states induced by its large spin-orbit coupling and its AFM transition below $T_N$ = 17 K [43]. First-principles calculations predict $EuIn_2As_2$ to host a long-awaited axion insulator state [43–45]. Theoretically, gapless Dirac surface states within the spin-orbit induced bulk band-gap in the paramagnetic phase open a surface state gap induced by the time reversal symmetry breaking in the AFM phase [44–48]. It may also host higher order TI with 1D chiral hinge state, depending on the orientation of magnetic moments [43–46]. Most importantly, DFT calculations reveal the topological surface states to reside right at the Fermi energy in its stoichiometric form, making it particularly interesting and distinct from the previously discovered material systems [43,45].

$EuIn_2As_2$ has an in-plane hexagonal structure with alternating Eu and In-As layers [49]. The $Eu^{2+}$ ions carry the $4f$ magnetic moment of $7\mu_B$ [47]. These localized electronic states reside ~1.7eV below the Fermi energy. Very recent neutron diffraction experiments find a broken helical AFM order in the bulk, where the magnetic moments order ferromagnetically within each Eu layer (ab-plane) and rotate to form a broken helix along the c-axis [44]. DFT calculations reveal that a sizable spin orbit coupling induces bulk band inversion of the As 4p hole-like states and the In 5s electron-like states, opening up a spin-orbit gap of ~100 meV at the Fermi energy that is robust and mostly insensitive to the magnetic order [43,45]. Within the spin-orbit gap reside the topologically non-trivial surface states. On the other hand, temperature dependent ARPES experiments show the system to be slightly hole doped, shifting the bands higher in energy and placing the spin-orbit gap just above the chemical potential, making the non-trivial surface states within it inaccessible to ARPES [45–48]. ARPES also observes another hole-like band analogous to the As 4p band that is vertically shifted and has no counterpart in the bulk or surface band structures and thus remains mysterious [48].

Here we use scanning tunneling microscopy and spectroscopy (STM/S) complemented with DFT calculations to probe the surface electronic states of cleaved single crystals of EuIn$_2$As$_2$. From our STM images of multiple cleaved single crystals, we find the cleaving to occur at the Eu layer. The cleaving rips the Eu layer apart exposing a partially covered Eu terminated surface. Through analyzing surfaces of multiple cleaved samples, we find that the percentage of the Eu atoms that remain on the surface varies locally with values ranging from less than 40% (a lower bound has not been determined) to almost 90%. Depending on the Eu surface density, the Eu ions undergo a surface reconstruction with either a 2a x 1a stripe order or an Ra x Ra bilayer atomic order, where a = 4.2Å is the in-plane lattice constant and R is commensurate and takes the values of 3/2 or $\sqrt{3}$ (see fig.1i). Our spectroscopic measurements (dI/dV) reveal multiple peaks within the density of states, which originate from the various Van Hove singularities attributed to the band edges/gaps of the surface and bulk band structures. Following the evolution of the electronic states across the AFM transition reveals the low temperature surface state peaks to move closer and partially overlap below T$_N$. Small yet a finite gap, however, remains above T$_N$.

Figure 1a-c show topographic images from three different cleaves of EuIn$_2$As$_2$ single crystals. Surfaces exposed from other cleaved crystals result in structures identical to these. All cleaves of the crystals have been carried out in-situ in ultra-high vacuum condition at room temperature and immediately transferred to the STM head that is held at 9.5 K. The cleaving of our samples at room temperature, as discussed below, results in different surfaces as compared to the recent STM work performed on cold cleaved samples [46]. The topographs reveal terraces separated by multiples of c/2, where c = 17.9 Å is the bulk out-of-plane lattice constant. Within terraces, however, we find surfaces with different surface structures separated by approximately 2-3 Å. All observed surfaces reveal structures that are different from the bulk indicating that they undergo a surface reconstruction. To identify the chemical nature of the surfaces, we examine the crystal structure of EuIn$_2$As$_2$ [50], which composes of [In$_2$As$_2$]$^{2-}$ and Eu$^{2+}$ layers [49] (Fig.1g), alternating along the crystallographic c-axis and balancing the net charge. Two possible cleaving planes can be identified from the crystal structure – breaking the In-In bond (bond length of 2.76 Å) and thus exposing the In-terminated surface or breaking down the Eu-layer (Eu-As (bond length of 3.10 Å) as well as Eu-Eu (bond length of 4.20 Å) bonds) and exposing a partially covered Eu-terminated surfaces [49]. Based on the bond lengths, the latter is more likely.

Indeed, our topographs provide evidence of the latter scenario only, exposing partially covered atomic surfaces, with varying atomic concentrations. Such structures will not be expected from In-In bond breaking, which should result in identical, fully covered, In-layers on both sides of the cleave. Depending on the Eu coverage left behind after the cleave, two different structures are formed on the surface. We find a 2a x 1a stripe order (where a = 4.2 Å is the bulk in-plane lattice constant) with three-fold symmetric domains (Fig.1d), indicating a 50% Eu surface coverage.

Reduced Eu-coverage beyond 50% disrupts the stripe order by introducing additional missing atoms (blue regions in Fig.1d). Further reduction leads to a disordered stripe structure (Fig.1e). On the other hand, for higher coverage exceeding 50%, the Eu atoms buckle up (along the c-axis) forming a double Eu-layer with a 1.5ax1.5a or $\sqrt{3}a \times \sqrt{3}a$ atomic structure with varying surface layer coverage, which we term atomic surface (Fig.1f).

Figure 2 shows a closer look at the atomic surface structure. When forming the 1.5a x 1.5a structure, the Eu atoms now require more surface space as they are farther apart, and thus only 44% of the atoms that would fit in one layer with a 1a x 1a structure can fit in this new reconstructed structure. Similarly, with the $\sqrt{3}a \times \sqrt{3}a$ structure, only 33% of the atoms would fit in one layer. The additional Eu atoms (beyond the 44% or the 33% depending on the formed structure) remaining on the surface after the cleave rearrange into a second layer on top of the base layer of Eu, thus forming a double layer of Eu atoms. A large-scale topography in Fig.2a shows the atomic structure in a double layer form. Analysis of the surface structure of this double layer confirms their identical reconstructed structure as shown in Fig.2b-d. The height difference between the bottom and top Eu atomic surfaces is about 1.5-2 Å. The Fourier transform of both (ordered and disordered) stripe surfaces and the atomic surface are shown in Fig.2e-g. Note that the stripe order Bragg peaks broaden with disorder (Fig.2e,f).

Spectroscopic dI/dV conductance measurements on the different surfaces at a temperature of 9.5 K (well below $T_N$ = 17 K) are shown in Figure 3. On the atomic surface (with full coverage), the spectra reveal three peaks within an energy range of 0 – 150 meV, whose positions slightly vary spatially (different colors in Fig.3a.). An additional hump can be seen just below $E_F$. Beyond 200 meV, the density of states mostly vanishes. This surface and corresponding spectra have not been seen in the previous STM study [46], presumable due to the warm (room temperature) cleaving in our experiments. On the other hand, similar to the previous STM work [46], on the stripe surface (Fig.3), besides the hump below $E_F$ and the similar three peaks seen in the 0 – 150 meV range, an additional set of peaks with similar energy spacing can be seen in the energy range of 150 – 300 meV.

Temperature dependence of the electronic states across the magnetic transition $T_N$ = 17K is shown in Figure 4a and Figure 5a for the stripe and atomic surfaces, respectively. Both surfaces show clear evolution of the density of states across $T_N$. Since the atomic surface has not been observed in the previous STM work, we focus our analysis on the atomic surface. On this surface (Fig.5a, b), the density of state peaks below $E_F$ and around 120 meV exhibit insignificant temperature dependence across $T_N$ (Fig.5b). On the other hand, the peaks between 10 and 60 meV move closer and partially merge with increasing temperature, leaving a small but finite gap above $T_N$ (Fig. 5b,c).

To understand the origin of the different peaks in the density of states, we carry out DFT calculations of the electronic band structure. We adopted the same lattice structure and constants used in previous DFT work and perform DFT+U calculations using WIEN2k package [51].

See methods section for the details of the DFT calculations. We reproduce the band structure in the AFM-c state (see Fig.6a). Note that the choice of magnetic order (AFM-a, AFM-c, AFM-helical) results in similar bulk band structure and spin orbit gap [43–45,48]. In accordance with previous reports [43,44], we find an electron-like band arising from the As $p_z$-orbitals and a hole-like band from the In s-orbitals near the Fermi energy that are inverted by the spin orbit coupling, opening a topological spin-orbit gap of ~100 meV at $E_F$. The DFT calculations also show the Eu $f$-states to be localized residing far below (~1.7eV) $E_F$ with no contribution to the density of states near $E_F$, in accordance with previous reports [43–45].

Fig.6b shows the band structure around the Gamma point near the Fermi energy. The solid green lines represent the bulk band structure. The bulk band edges, just below $E_F$ and around 90-120 meV, represent Van Hove singularities that lead to three peaks labelled B1, B2 and B3 in the calculated density of states (Fig.6c). These peak energies nicely agree with the atomic-surface STM spectra of the hump below $E_F$ and the peak near ~120 meV (Fig.3a, Fig.5). Note that the corresponding B3 peak (~120 meV) seen in the experiment is dominant as compared to the B1 and B2. This is expected by the fact that the STM tunneling to the As-layer through the surface Eu-layer dominates as compared to the tunneling to the In-layer located further below the surface. The DFT results (for the band below $E_F$) also well agree with the recent ARPES data of the bulk band structure [45].

Motivated by the recent predictions of the gapped surface states on EuIn$_2$As$_2$ [44], we model the surface band structures in the magnetic states with the following dispersion, $E^s(\vec{k},\theta) = \mu^s \pm \sqrt{\varepsilon^2(\vec{k}) + m^2(\vec{k})}$, where $\vec{k}$ is the 2D momentum measured from the $\Gamma$ point, reflecting the 2D nature of the surface states. $\varepsilon(k)$ is the surface state dispersion without a gap, which can be generally written as $\varepsilon(k) = ak + bk^2 + ck^3$. $\mu^s$ is the location in energy of the Dirac point, which is chosen to be 43 meV in order to fix the Dirac point at the center of the bulk gap. The $k$-dependent gap function $m(\vec{k})$ is introduced because of the anisotropy observed in the previous studies [44]. It should be noted that $m(\vec{k})$ is the gap of the surface state which is different from the bulk gap and is sensitive to the surface potential. While it is not possible to know the surface potential to obtain the realistic surface state gap, it is expected that $m(k)$ could be highly inhomogeneous so that it has strong a $k$-dependence. By examining our STM data, we find that the gap function with a nodal line is crucial to reproduce the line shape of the surface state DOS. The simplest choice is $m(\vec{k}) = \frac{\Delta}{2}(1 - sin\theta)$.

With the surface state dispersion, we are able to compute the surface state DOS by evaluating

$$D^s(E) = \alpha \int_0^{k_{max}} dk\, k \int_0^{2\pi} d\theta \delta(E - E(\vec{k}))$$

where $\alpha$ is the factor modulating the ratio between surface state DOS and bulk DOS, and $k_{max}$ is the maximal momentum at which the surface state can exist. The final DOS presented in the main text is obtained by

$$D(E) = D^B(E) + D^s(E)$$

where $D^B(E)$ is the bulk DOS obtained directly from the DFT calculations, and parameters of $ak_{max} = 10$ meV, $bk_{max}^2 = 12$ meV, $ck_{max}^3 = 28$ meV, and $\Delta = 12.5$ meV are used in $D^s(E)$.

The resulting surface band structure is shown as the dashed red lines in Fig.6b. Here we want to emphasize that the surface band structure only represents a simple model that captures the overall structure seen in the most recent surface state DFT calculations. However, the fine details of such surface state calculations strongly depend on the magnetic ordering on the terminated Eu-surface, which experimentally is quite challenging to probe. Fortunately, however, recent DFT calculations show that the topological gapless surface states and Axion insulator are expected for both in-plane as well as helical magnetic orders [44,47,52], thus maintaining the same topological phase. Regardless of the fine details, the modeled gapped surface band structure leads to additional two peaks, labeled S1 and S2, located within the bulk spin-orbit gap, that emerge due to the partially gapped surface state band edges (see Fig.6b and 6c).

The overall calculated DOS is in good agreement with the STM spectra measured on the atomic surface (Fig.3a and Fig.5). From the temperature dependent spectra shown in Fig.5, we can attribute the hump at $E_F$ and the peak at 120 meV to the B1 and B3 bulk valence and conduction band edges. Fitting these peaks to a Gaussian function, we find the peak positions to remain mostly unchanged across the magnetic transition. As these peaks correspond to the spin-orbit induced bulk gap, they are not expected to be strongly affected by the magnetic order as has been seen in DFT [45]. On the other hand, the origin of the peaks around 10 and 60 meV at low temperatures can be attributed to the topological surface state band edges S1 and S2 (see Fig.5 and 6). With increasing temperature these surface state bands evolve rapidly and move closer. In particular, the S2 surface state drops dramatically and within its bandwidth (of 20 meV) partially merges with the S1 surface band above $T_N$ (Fig. 5b). Fitting these peaks to Gaussians, one can see a clear evolution of the S1/S2 peaks (Fig.5b, c). While the two bands, with their bandwidths of about 20 meV extracted from the Gaussian fits, mostly overlap, the peak energies remain slightly separate above $T_N$, which is better seen in the fits (Fig.5c). We want to emphasize that while a surface state gap structure can be deduced from our data in the AFM phase, the model shown here is not unique. For example, an indirect gap structure, similar to what has been calculated in recent DFT work for in-plane and out of plane magnetic order [45] can also reproduce an analogous density of state profile. Therefore, regardless of the fine detail, the STM data provide evidence of a topological surface state partial gap of ~40 meV at low temperature in the AFM phase, residing within a ~120 meV spin-orbit induced bulk gap. With increasing temperature, the gap strongly decreases but remains finite above $T_N$.

Our DFT calculations also provide an explanation of the spectra on the stripe surface. The stripe order introduces additional bands within the surface electronic band structure. The STM data suggests that in the stripe-ordered surface, one of every two Eu atoms on the surface is missing (or possibly closer to the As plane), which produce extra local potential on the electronic orbitals on the As atom nearby (see Fig.7a and Fig.8a). This effect can be reasonably modelled as a chemical potential of $\mu(\vec{R}) = E_F + Ve^{i\vec{Q}\cdot\vec{R}}$, where $\vec{Q}$ is the wave vector corresponding to the new periodicity induced by the ordering of Eu atoms. The tight-binding model with the site-dependent chemical potential in the momentum space can be written as

$$\hat{H} = \hat{H}(\vec{k}) + \hat{H}(\vec{k}+\vec{Q}) + \hat{H}_c(\vec{k},\vec{k}+\vec{Q})$$

where $\hat{H}_c(\vec{k},\vec{k}+\vec{Q})$ describes the Hamiltonian coupling $\vec{k}$ and $\vec{k}+\vec{Q}$ via the site-dependent chemical potential. As we can see from the band structure (Fig.6), there is only one band in the energy widow between -200meV and +200meV, thus the Hamiltonian in this range of energy can be approximated as a two-level system in the form of

$$\begin{pmatrix} \varepsilon(\vec{k}) & V \\ V & \varepsilon(\vec{k}+\vec{Q}) \end{pmatrix}$$

where $\varepsilon(\vec{k})$ is the energy dispersion of the band in the energy widow between -200meV and +200meV. Diagonalizing the above effective Hamiltonian gives eigen energies of

$$E_\pm(\vec{k}) = \frac{\varepsilon(\vec{k})+\varepsilon(\vec{k}+\vec{Q})}{2} \pm \sqrt{\left(\frac{\varepsilon(\vec{k})-\varepsilon(\vec{k}+\vec{Q})}{2}\right)^2 + V^2}.$$

The nature of the new bands $E_\pm(k)$ crucially depends on the energy difference of $(\varepsilon(k) - \varepsilon(k+Q))$. If the difference is large compared to $V$, the new bands become

$$E_\pm(\vec{k}) \approx \frac{\varepsilon(\vec{k})+\varepsilon(\vec{k}+\vec{Q})}{2} \pm \frac{|\varepsilon(\vec{k})-\varepsilon(\vec{k}+\vec{Q})|}{2} \pm O(V^2).$$

In this limit, the effective Hamiltonian essentially gives the same bands $\varepsilon(\vec{k})$ and $\varepsilon(\vec{k}+\vec{Q})$ up to corrections of order of $V^2$, and no band replication will be observed. In the case of the atomic ordering (Ra x Ra with R being 3/2 or $\sqrt{3}$), the corresponding ordering wave vector $\vec{Q}$ lies between $\Gamma$ and $M$ points, and it can be seen from the band structure that $\varepsilon(\vec{k}+\vec{Q}) > \varepsilon(\vec{k})$ for all the $\vec{k}$ points near $\Gamma$ point. As a result, no band replication will be seen at energy near the Fermi energy.

On the other hand, if the difference is small compared to $V$, the new bands become

$$E_\pm(\vec{k}) \approx \frac{\varepsilon(\vec{k}) + \varepsilon(\vec{k}+\vec{Q})}{2} \pm |V| \pm O(\epsilon^2) \approx \varepsilon(\vec{k}) \pm |V| + \frac{\epsilon}{2} \pm O(\epsilon^2)$$

where $\epsilon \equiv \varepsilon(\vec{k}) - \varepsilon(\vec{k} + \vec{Q})$ is a small parameter now. In this limit, since $\epsilon \to 0$, $E_\pm(\vec{k}) \approx \varepsilon(\vec{k}) \pm |V|$, which results in the band replication near the $\Gamma$ point at low energy around the Fermi energy. For the case of stripe-ordering Eu atom, $\vec{Q} = (0.5,0,0) = \vec{X}$ and it can be seen from the band structure that $\varepsilon(\vec{k} + \vec{Q}) \approx \varepsilon(\vec{k})$ for all the $\vec{k}$ points near $\Gamma$ point. As a result, the band replication can be observed, and the band splitting is controlled by $V$ (see Fig.7b and Fig.8b).

The replication of the band structure due to the striped ordered of Eu atom is quite general. In principle, since the stripe order of Eu atom emerges only near the surface, a DFT calculation of a semi-infinite system terminated by the stripe-ordered surface is necessary but unfortunately not possible. To demonstrate the physics, we make two artificial structures that resemble the stripe order but still have some crystal symmetries making DFT calculations feasible. These two artificial structures could be seen as two limiting cases for the degrees of the band replication (Fig.7, 8).

The amount of energy-shift of the replica band depends on the magnitude of the chemical potential modulation. The STM spectra on the ordered/disordered stripe surfaces (Fig.3b,c) can nicely be accounted for by this scenario. The replication of the bands thus explains the additional higher energy (150-300 meV) set of peaks seen in Fig.3b,c, which do appear as replica of the lower energy (0-150 meV) peaks. This finding also provides a natural explanation of the mysterious surface state replica band seen in recent ARPES studies (i.e, see Fig.3 in ref. 48). We also note that replica bands also form on the atomic surface. However, due to their different reconstruction wave-vector, the bands lie away from the interesting region near the Fermi energy.

In conclusion, our data, regardless of the fine details, point to a topological gap induced by magnetic order and suggests that an axion insulator state may be possible to realize in $EuIn_2As_2$. While $EuIn_2As_2$ in its single crystalline form displays finite bulk and surface conductivity and is not truly an insulator, an insulating state may be realized by doping, strain, or surface engineering. In fact, epitaxial films on a proper substrate may solve these issues. The STM data here indicate that the bulk gap is located just above $E_F$, with its band edge crossing the Fermi energy by only a few meV. On the other hand, the surface states are additionally hole doped and replicated, induced by the missing Eu atoms. The surface gap is thus located only ~30 meV above $E_F$. Therefore, minute perturbations may be enough to drive epitaxial $EuIn_2As_2$ from a semi-metallic into a true insulating phase and thus realizing the anticipated axion state.

## Appendix 1: Single crystal growth

Single crystals of $EuIn_2As_2$ were grown from In-As flux. High purity constituent elements (purity: Eu 3N, In 5N, As 5N) were taken in a molar ratio Eu : In : As of 1:12:3. The synthesis was carried out using a ACP-CCS-5 Canfield Crucible Set (LCP Industrial Ceramics Inc.) sealed in an evacuated quartz tube. The mixture was heated up to 1000 ºC at a rate of 50 ºC/h, maintained at this temperature for 24 hours, and then slowly cooled down to 700 ºC at a rate 2 ºC/h. Subsequently, the ampule was removed from the furnace, flipped over and centrifuged in order to remove the flux mixture. As a product of the entire procedure, several silver-shiny platelet-like crystals were obtained with dimensions up to 2 x 2 x 0.1 mm. They were found stable against air and moisture.

Chemical composition and phase homogeneity of the prepared crystals were determined by energy-dispersive X-ray (EDX) analysis performed using a FEI scanning electron microscope equipped with an EDAX Genesis XM4 spectrometer. In order to verify the crystal symmetry of the obtained single crystals, a small fragment was crumbled from a larger piece and examined on an Oxford Diffraction X'calibur four-circle single-crystal X-ray diffractometer equipped with a CCD Atlas detector. Crystallinity and crystallographic orientation of the particular crystals used in physical measurements were determined by means of Laue X-ray backscattering technique implemented in a LAUE-COS (Proto) system.

## Appendix 2: Details of DFT calculations

The DFT+U calculations are carried out with full potential linear augmented plane waves plus local orbitals (FP-LAPW+lo) and the Perdew-Burke-Ernzerhof generalized gradient approximation (PBE-GGA) provided in the WIEN2k code [53,54]. The Hubbard U terms are treated by the approach of the self-interaction corrections developed by Anisimov et al. [55], which is available as SIC scheme in WIEN2k package.

We have used the same crystal structure and lattice constants provided by Y. Xu et. al. [43] and a k-mesh of 38X38X7 is used to sample the Brillouin zone. The value of U is chosen to be 5.0 eV with J fixed to be J=0.2U. All the DFT+U calculations are done with the inclusion of spin-orbit coupling on every atom.

The tight-binding model employed for the calculation of band structures and DOS composed of the s and f orbitals of Eu atom, the s and p orbitals of In atom, and the s and p orbitals of As atom. Hopping parameters are determined by fitting the DFT band structure using the Wannier90 [56].

**Acknowledgments**

PA acknowledges funding from the U.S. National Science Foundation (NSF) CAREER under award No. DMR-1654482. DK was supported by the National Science Centre (Poland) under research grant 2021/41/B/ST3/01141.

*These authors contributed equally to this work.
†aynajian@binghamton.edu,
‡wlee@binghamton.edu

[1]   B. A. Bernevig, C. Felser, and H. Beidenkopf, *Progress and Prospects in Magnetic Topological Materials*, Nature **603**, 41 (2022).

[2]   M. Ezawa, *Magnetic Second-Order Topological Insulators and Semimetals*, Phys. Rev. B **97**, 155305 (2018).

[3]   R. X. Zhang, F. Wu, and S. Das Sarma, *Möbius Insulator and Higher-Order Topology in $MnBi_{2n}Te_{3n+1}$*, Phys. Rev. Lett. **124**, 136407 (2020).

[4]   I. Belopolski, K. Manna, D. S. Sanchez, G. Chang, B. Ernst, J. Yin, S. S. Zhang, T. Cochran, N. Shumiya, H. Zheng, et al., *Discovery of Topological Weyl Fermion Lines and Drumhead Surface States in a Room Temperature Magnet*, Science **365**, 1278 (2019).

[5]   D. F. Liu, A. J. Liang, E. K. Liu, Q. N. Xu, Y. W. Li, C. Chen, D. Pei, W. J. Shi, S. K. Mo, P. Dudin, et al., *Magnetic Weyl Semimetal Phase in a Kagomé Crystal*, Science **365**, 1282 (2019).

[6]   Y. Tokura, K. Yasuda, and A. Tsukazaki, *Magnetic Topological Insulators*, Nat. Rev. Phys. **1**, 126 (2019).

[7]   P. Wang, J. Ge, J. Li, Y. Liu, Y. Xu, and J. Wang, *Intrinsic Magnetic Topological Insulators*, Innov. **2**, 100098 (2021).

[8]   E. H. da Silva Neto, *"Weyl"Ing Away Time-Reversal Symmetry*, Science **365**, 1248 (2019).

[9]   C. X. Liu, S. C. Zhang, and X. L. Qi, *The Quantum Anomalous Hall Effect: Theory and Experiment*, Annu. Rev. Condens. Matter Phys. **7**, 301 (2016).

[10]  Y. Deng, Y. Yu, M. Z. Shi, Z. Guo, Z. Xu, J. Wang, X. H. Chen, and Y. Zhang, *Quantum Anomalous Hall Effect in Intrinsic Magnetic Topological Insulator $MnBi_2Te_4$*, Science **367**, 895 (2020).

[11]  Q. L. He, L. Pan, A. L. Stern, E. C. Burks, X. Che, G. Yin, J. Wang, B. Lian, Q. Zhou, E. S. Choi, et al., *Chiral Majorana Fermion Modes in a Quantum Anomalous Hall Insulator–Superconductor Structure*, Science **357**, 294 (2017).

[12]  J. G. Checkelsky, R. Yoshimi, A. Tsukazaki, K. S. Takahashi, Y. Kozuka, J. Falson, M. Kawasaki, and Y. Tokura, *Trajectory of the Anomalous Hall Effect towards the Quantized State in a Ferromagnetic Topological Insulator*, Nat. Phys. **10**, 731 (2014).

[13]  C. Liu, Y. Wang, H. Li, Y. Wu, Y. Li, J. Li, K. He, Y. Xu, J. Zhang, and Y. Wang, *Robust Axion Insulator and Chern Insulator Phases in a Two-Dimensional Antiferromagnetic Topological Insulator*, Nat. Mater. **19**, 522 (2020).

[14]  R. Li, J. Wang, X. L. Qi, and S. C. Zhang, *Dynamical Axion Field in Topological Magnetic*

*Insulators*, Nat. Phys. **6**, 284 (2010).

[15] S. Nadj-Perge, I. K. Drozdov, J. Li, H. Chen, S. Jeon, J. Seo, A. H. MacDonald, B. A. Bernevig, and A. Yazdani, *Observation of Majorana Fermions in Ferromagnetic Atomic Chains on a Superconductor*, Science **346**, 602 (2014).

[16] N. C. Frey, M. K. Horton, J. M. Munro, S. M. Griffin, K. A. Persson, and V. B. Shenoy, *High-Throughput Search for Magnetic and Topological Order in Transition Metal Oxides*, Sci. Adv. **6**, 702 (2020).

[17] Y. Xu, L. Elcoro, Z. Da Song, B. J. Wieder, M. G. Vergniory, N. Regnault, Y. Chen, C. Felser, and B. A. Bernevig, *High-Throughput Calculations of Magnetic Topological Materials*, Nature **586**, 702 (2020).

[18] L. Elcoro, B. J. Wieder, Z. Song, Y. Xu, B. Bradlyn, and B. A. Bernevig, *Magnetic Topological Quantum Chemistry*, Nat. Commun. **12**, 5965 (2021).

[19] M. M. Otrokov, I. I. Klimovskikh, H. Bentmann, D. Estyunin, A. Zeugner, Z. S. Aliev, S. Gaß, A. U.B. Wolter, A. V. Koroleva, A. M. Shikin, et al., *Prediction and Observation of an Antiferromagnetic Topological Insulator*, Nature **576**, 416 (2019).

[20] S. Lei, A. Saltzman, and L. M. Schoop, *Complex Magnetic Phases Enriched by Charge Density Waves in the Topological Semimetals $GdSb_xTe_{2-x-\delta}$*, Phys. Rev. B **103**, 134418 (2021).

[21] N. B. M. Schröter, I. Robredo, S. Klemenz, R. J. Kirby, J. A. Krieger, D. Pei, T. Yu, S. Stolz, T. Schmitt, P. Dudin, et al., *Weyl Fermions, Fermi Arcs, and Minority-Spin Carriers in Ferromagnetic $CoS_2$*, Sci. Adv. **6**, 5000 (2020).

[22] A. Noah, F. Toric, T. D. Feld, G. Zissman, A. Gutfreund, D. Tsruya, T. R. Devidas, H. Alpern, A. Vakahi, H. Steinberg, et al., *Tunable Exchange Bias in the Magnetic Weyl Semimetal $Co_3Sn_2S_2$*, Phys. Rev. B **105**, 144423 (2022).

[23] G. M. Pierantozzi, A. De Vita, C. Bigi, X. Gui, H. J. Tien, D. Mondal, F. Mazzola, J. Fujii, I. Vobornik, G. Vinai, et al., *Evidence of Magnetism-Induced Topological Protection in the Axion Insulator Candidate $EuSn_2P_2$*, Proc. Natl. Acad. Sci. U. S. A. **119**, e2116575119 (2022).

[24] E. Liu, Y. Sun, N. Kumar, L. Muechler, A. Sun, L. Jiao, S. Y. Yang, D. Liu, A. Liang, Q. Xu, et al., *Giant Anomalous Hall Effect in a Ferromagnetic Kagome-Lattice Semimetal*, Nat. Phys. **14**, 1125 (2018).

[25] A. Sakai, Y. P. Mizuta, A. A. Nugroho, R. Sihombing, T. Koretsune, M. T. Suzuki, N. Takemori, R. Ishii, D. Nishio-Hamane, R. Arita, P. Goswami, and S. Nakatsuji, *Giant Anomalous Nernst Effect and Quantum-Critical Scaling in a Ferromagnetic Semimetal*, Nat. Phys. **14**, 1119 (2018).

[26] P. Li, J. Koo, W. Ning, J. Li, L. Miao, L. Min, Y. Zhu, Y. Wang, N. Alem, C. X. Liu, Z. Mao, and B. Yan, *Giant Room Temperature Anomalous Hall Effect and Tunable Topology in a*

*Ferromagnetic Topological Semimetal $Co_2MnAl$*, Nat. Commun. **11**, 3476 (2020).

[27] M. Hirschberger, S. Kushwaha, Z. Wang, Q. Gibson, S. Liang, C. A. Belvin, B. A. Bernevig, R. J. Cava, and N. P. Ong, *The Chiral Anomaly and Thermopower of Weyl Fermions in the Half-Heusler GdPtBi*, Nat. Mater. **15**, 1161 (2016).

[28] S. Borisenko, D. Evtushinsky, Q. Gibson, A. Yaresko, K. Koepernik, T. Kim, M. Ali, J. van den Brink, M. Hoesch, A. Fedorov, et al., *Time-Reversal Symmetry Breaking Type-II Weyl State in $YbMnBi_2$*, Nat. Commun. **10**, 3424 (2019).

[29] I. Lee, C. K. Kim, J. Lee, S. J.L. Billinge, R. Zhong, J. A. Schneeloch, T. Liu, T. Valla, J. M. Tranquada, G. Gu, and J. C.S. Davis, *Imaging Dirac-Mass Disorder from Magnetic Dopant Atoms in the Ferromagnetic Topological Insulator $Cr_x(Bi_{0.1}Sb_{0.9})_{2-x}Te_3$*, Proc. Natl. Acad. Sci. U. S. A. **112**, 1316 (2015).

[30] X. L. Qi, R. Li, J. Zang, and S. C. Zhang, *Inducing a Magnetic Monopole with Topological Surface States*, Science **323**, 1184 (2009).

[31] C. Nayak, S. H. Simon, A. Stern, M. Freedman, and S. Das Sarma, *Non-Abelian Anyons and Topological Quantum Computation*, Rev. Mod. Phys. **80**, 1083 (2008).

[32] S. Howard, L. Jiao, Z. Wang, N. Morali, R. Batabyal, P. Kumar-Nag, N. Avraham, H. Beidenkopf, P. Vir, E. Liu, et al., *Evidence for One-Dimensional Chiral Edge States in a Magnetic Weyl Semimetal $Co_3Sn_2S_2$*, Nat. Commun. **12**, 4269 (2021).

[33] C. Yue, Y. Xu, Z. Song, H. Weng, Y. M. Lu, C. Fang, and X. Dai, *Symmetry-Enforced Chiral Hinge States and Surface Quantum Anomalous Hall Effect in the Magnetic Axion Insulator $Bi_{2-x}Sm_xSe_3$*, Nat. Phys. **15**, 577 (2019).

[34] J. Teng, N. Liu, and Y. Li, *Mn-Doped Topological Insulators: A Review*, J. Semicond. **40**, 081507 (2019).

[35] M. Mogi, M. Kawamura, R. Yoshimi, A. Tsukazaki, Y. Kozuka, N. Shirakawa, K. S. Takahashi, M. Kawasaki, and Y. Tokura, *A Magnetic Heterostructure of Topological Insulators as a Candidate for an Axion Insulator*, Nat. Mater. **16**, 516 (2017).

[36] D. Xiao, J. Jiang, J. H. Shin, W. Wang, F. Wang, Y. F. Zhao, C. Liu, W. Wu, M. H.W. Chan, N. Samarth, and C. Z. Chang, *Realization of the Axion Insulator State in Quantum Anomalous Hall Sandwich Heterostructures*, Phys. Rev. Lett. **120**, 056801 (2018).

[37] K. He, *$MnBi_2Te_4$-Family Intrinsic Magnetic Topological Materials*, Npj Quantum Mater. **5**, 90 (2020).

[38] H. Li, S. Y. Gao, S. F. Duan, Y. F. Xu, K. J. Zhu, S. J. Tian, J. C. Gao, W. H. Fan, Z. C. Rao, J. R. Huang, et al., *Dirac Surface States in Intrinsic Magnetic Topological Insulators $EuSn_2As_2$ and $MnBi_{2n}Te_{3n+1}$*, Phys. Rev. X **91**, 041039 (2019).

[39] S. H. Lee, Y. Zhu, Y. Wang, L. Miao, T. Pillsbury, H. Yi, S. Kempinger, J. Hu, C. A. Heikes, P. Quarterman, et al., *Spin Scattering and Noncollinear Spin Structure-Induced Intrinsic Anomalous Hall Effect in Antiferromagnetic Topological Insulator $MnBi_2Te_4$*, Phys. Rev.

Res. **1**, 012011 (2019).

[40] A. M. Shikin, D. A. Estyunin, I. I. Klimovskikh, S. O. Filnov, E. F. Schwier, S. Kumar, K. Miyamoto, T. Okuda, A. Kimura, K. Kuroda, et al., *Nature of the Dirac Gap Modulation and Surface Magnetic Interaction in Axion Antiferromagnetic Topological Insulator MnBi$_2$Te$_4$*, Sci. Rep. **10**, 13226 (2020).

[41] B. Chen, F. Fei, D. Zhang, B. Zhang, W. Liu, S. Zhang, P. Wang, B. Wei, Y. Zhang, Z. Zuo, et al., *Intrinsic Magnetic Topological Insulator Phases in the Sb Doped MnBi$_2$Te$_4$ Bulks and Thin Flakes*, Nat. Commun. **10**, 4469 (2019).

[42] C. Yan, S. Fernandez-Mulligan, R. Mei, S. H. Lee, N. Protic, R. Fukumori, B. Yan, C. Liu, Z. Mao, and S. Yang, *Origins of Electronic Bands in the Antiferromagnetic Topological Insulator MnBi$_2$Te$_4$*, Phys. Rev. B **104**, L041102 (2021).

[43] Y. Xu, Z. Song, Z. Wang, H. Weng, and X. Dai, *Higher-Order Topology of the Axion Insulator EuIn$_2$As$_2$*, Phys. Rev. Lett. **122**, 256402 (2019).

[44] S. X. M. Riberolles, T. V. Trevisan, B. Kuthanazhi, T. W. Heitmann, F. Ye, D. C. Johnston, S. L. Bud'ko, D. H. Ryan, P. C. Canfield, A. Kreyssig, et al., *Magnetic Crystalline-Symmetry-Protected Axion Electrodynamics and Field-Tunable Unpinned Dirac Cones in EuIn$_2$As$_2$*, Nat. Commun. **12**, 999 (2021).

[45] S. Regmi, M. M. Hosen, B. Ghosh, B. Singh, G. Dhakal, C. Sims, B. Wang, F. Kabir, K. Dimitri, Y. Liu, et al., *Temperature-Dependent Electronic Structure in a Higher-Order Topological Insulator Candidate EuIn$_2$As$_2$*, Phys. Rev. B **102**, 165153 (2020).

[46] Y. Li, H. B. Deng, C. X. Wang, S. S. Li, L. M. Liu, C. J. Zhu, K. Jia, Y. K. Sun, X. Du, X. Yu, et al., *Surface and Electronic Structure of Antiferromagnetic Axion Insulator Candidate EuIn$_2$As$_2$*, Wuli Xuebao/Acta Phys. Sin. **70**, 186801 (2021).

[47] Y. Zhang, K. Deng, X. Zhang, M. Wang, Y. Wang, C. Liu, J. W. Mei, S. Kumar, E. F. Schwier, K. Shimada, et al., *In-Plane Antiferromagnetic Moments and Magnetic Polaron in the Axion Topological Insulator Candidate EuIn$_2$As$_2$*, Phys. Rev. B **101**, 205126 (2020).

[48] T. Sato, Z. Wang, D. Takane, S. Souma, C. Cui, Y. Li, K. Nakayama, T. Kawakami, Y. Kubota, C. Cacho, et al., *Signature of Band Inversion in the Antiferromagnetic Phase of Axion Insulator Candidate EuIn$_2$As$_2$*, Phys. Rev. Res. **2**, 033342 (2020).

[49] A. M. Goforth, P. Klavins, J. C. Fettinger, and S. M. Kauzlarich, *Magnetic Properties and Negative Colossal Magnetoresistance of the Rare Earth Zintl Phase EuIn$_2$As$_2$*, Inorg. Chem. **47**, 11048 (2008).

[50] K. Momma and F. Izumi, *VESTA 3 for Three-Dimensional Visualization of Crystal, Volumetric and Morphology Data*, J. Appl. Crystallogr. **44**, 1272 (2011).

[51] P. Blaha, K. Schwarz, P. Sorantin, and S. B. Trickey, *Full-Potential, Linearized Augmented Plane Wave Programs for Crystalline Systems*, Comput. Phys. Commun. **59**, 399 (1990).

[52] C. Fang and L. Fu, *New Classes of Three-Dimensional Topological Crystalline Insulators:*

*Nonsymmorphic and Magnetic*, Phys. Rev. B - Condens. Matter Mater. Phys. **91**, 161105 (2015).

[53] K. Schwarz and P. Blaha, *Solid State Calculations Using WIEN2k*, Comput. Mater. Sci. **28**, 259 (2003).

[54] P. Blaha, K. Schwarz, F. Tran, R. Laskowski, G. K. H. Madsen, and L. D. Marks, *WIEN2k: An APW+lo Program for Calculating the Properties of Solids*, J. Chem. Phys. **152**, (2020).

[55] V. I. Anisimov, I. V. Solovyev, M. A. Korotin, M. T. Czyyk, and G. A. Sawatzky, *Density-Functional Theory and NiO Photoemission Spectra*, Phys. Rev. B **48**, 16929 (1993).

[56] A. A. Mostofi, J. R. Yates, Y. S. Lee, I. Souza, D. Vanderbilt, and N. Marzari, *Wannier90: A Tool for Obtaining Maximally-Localised Wannier Functions*, Comput. Phys. Commun. **178**, 685 (2008).

**Figures**

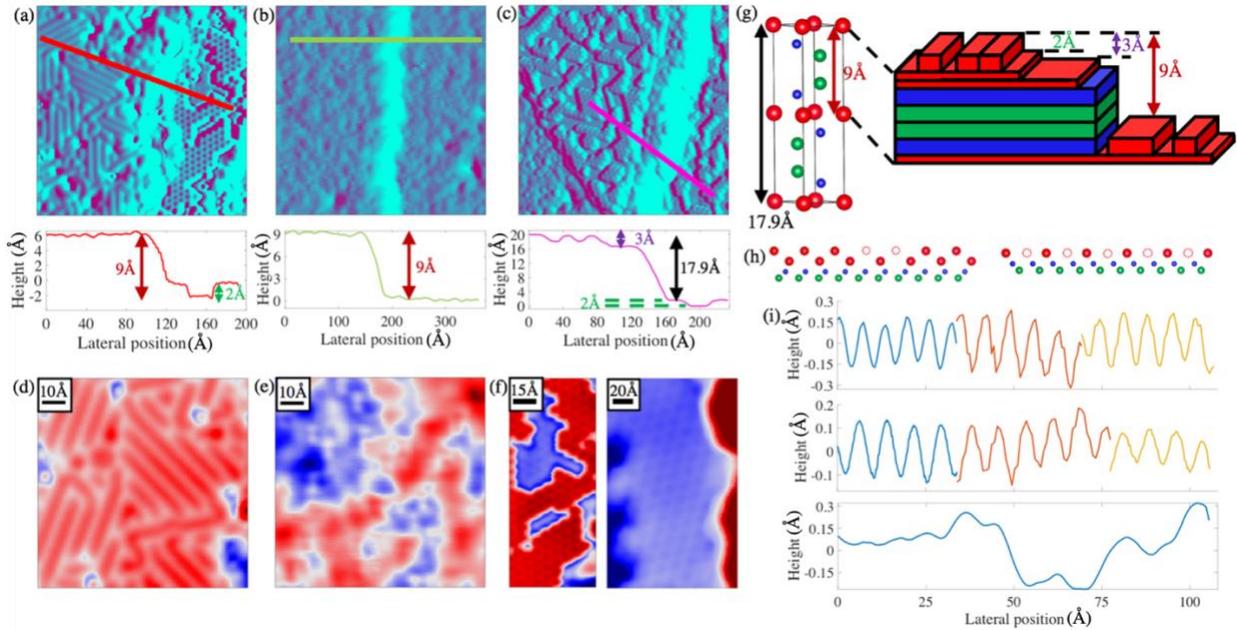

Figure 1. Topographic image of cleaved EuIn$_2$As$_2$ single crystal showing (a) the stripe and atomic surfaces (-200mV, -30pA), (b) the disordered stripe surface (-200mV, -30pA) and (c) large scale atomic surface with the disordered stripe surface (-200mV, -300pA). The lower insets correspond to the topographic height across the lines in (a, b, c). Zoom-in of the surface structure of the (d) the stripe surface (-200mV, -100pA), (e) the disordered stripe surface (-200mV, -100pA). (f) the atomic surface with different surface coverage (left: -200mV, -30pA; right: 200mV, -150pA). (g) Crystal structure of EuIn$_2$As$_2$ and a corresponding schematic of the layered structure showing the different observed cleaves with distances between different levels indicated (red = Eu, blue = As, green = In). (h) Schematics of the two observed reconstructions. The schematic on the left represents the atomic surface, while the one on the right represents the stripe surface. (i) Topographic line-cuts across the atomic, stripe, and disordered stripe surfaces. The different colors correspond to the three different directions.

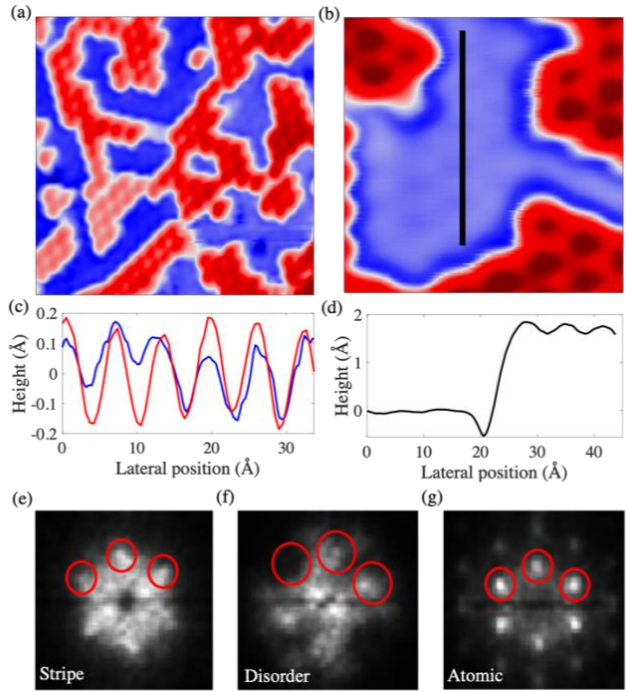

Figure 2. (a) Topography (-200meV,-300pA) showing the large scale atomic structure exhibiting two layers. (b) Close-in topography (-200meV, -50pA) of the double atomic layer, with good visibility of the structure within the lower layer (blue surface). (c) Two linecuts. Blue line is the linecut along the black line shown in (b), going along the atomic structure seen within the lower atomic layer (with the amplitude multiplied by 3.5). The calculated lattice constant is 6.45Å, which agrees strongly with the 1.5a reconstruction of the most commonly seen upper layer atomic surface. The red line is a linecut along the upper atomic surface, showing the same periodicity. (d) Linecut showing the step between the lower and upper atomic layers, with a step height of around 2Å. (e-g) Fourier transforms of the stripe, disordered stripe, and atomic surfaces. Red circles correspond to the Bragg peaks of the reconstructed surfaces.

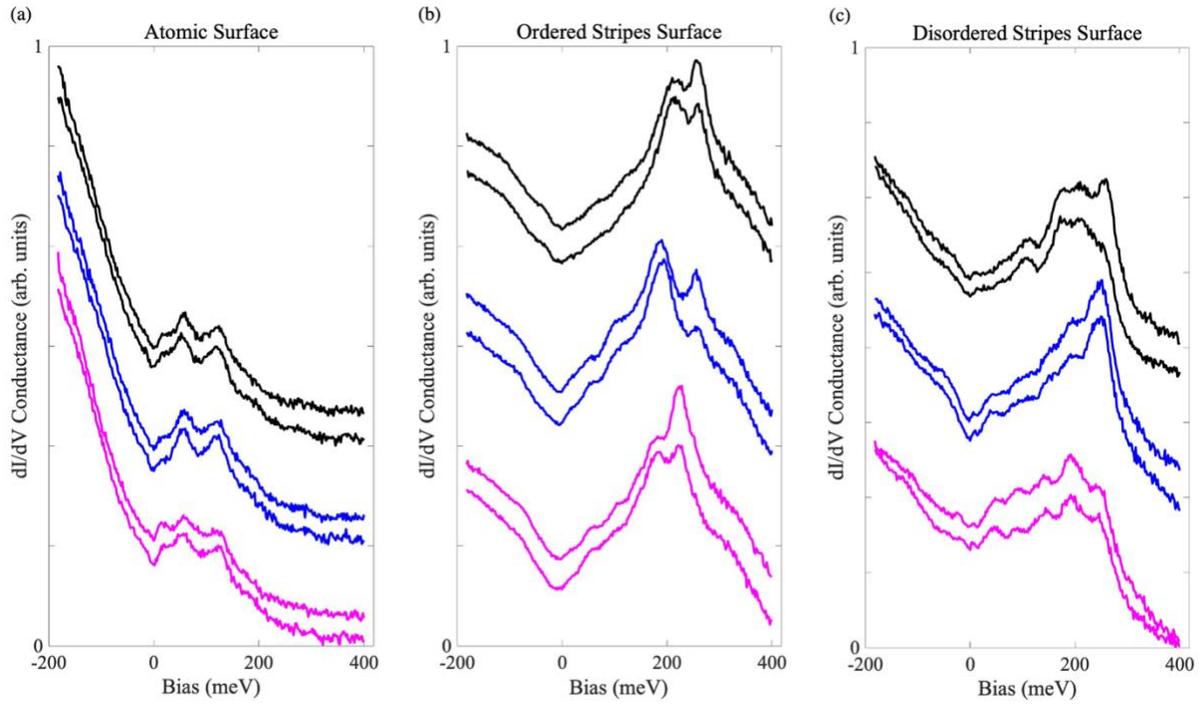

Figure 3. Spatial dependence of the conductance (dI/dV) at varying locations (shown above by the three different colors in each panel) on the three different surfaces: (a) atomic, (b) ordered and (c) disordered stripes. Same (different) color corresponds to near-by (further away) pixel positions where the spectra are acquired. All measurements are carried out at 9.5K with setpoint bias of -200 meV and a setpoint current of 100 pA.

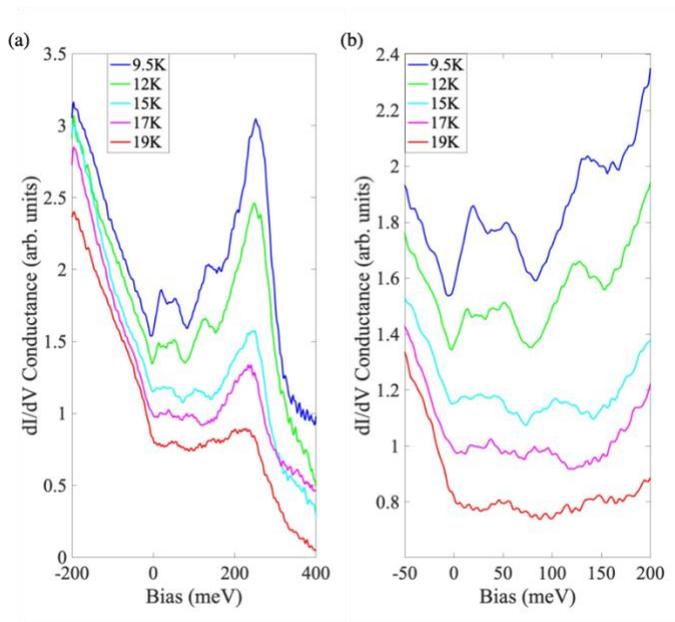

Figure 4. Temperature-dependence of dI/dV conductance spectra on the disordered stripe surface with energy ranges of (a) -200 to 400 meV and (b) -50 to 250 meV. All spectra are taken at the exact same location on the disordered stripe surface with bias = -200meV and setpoint current = -100mA.

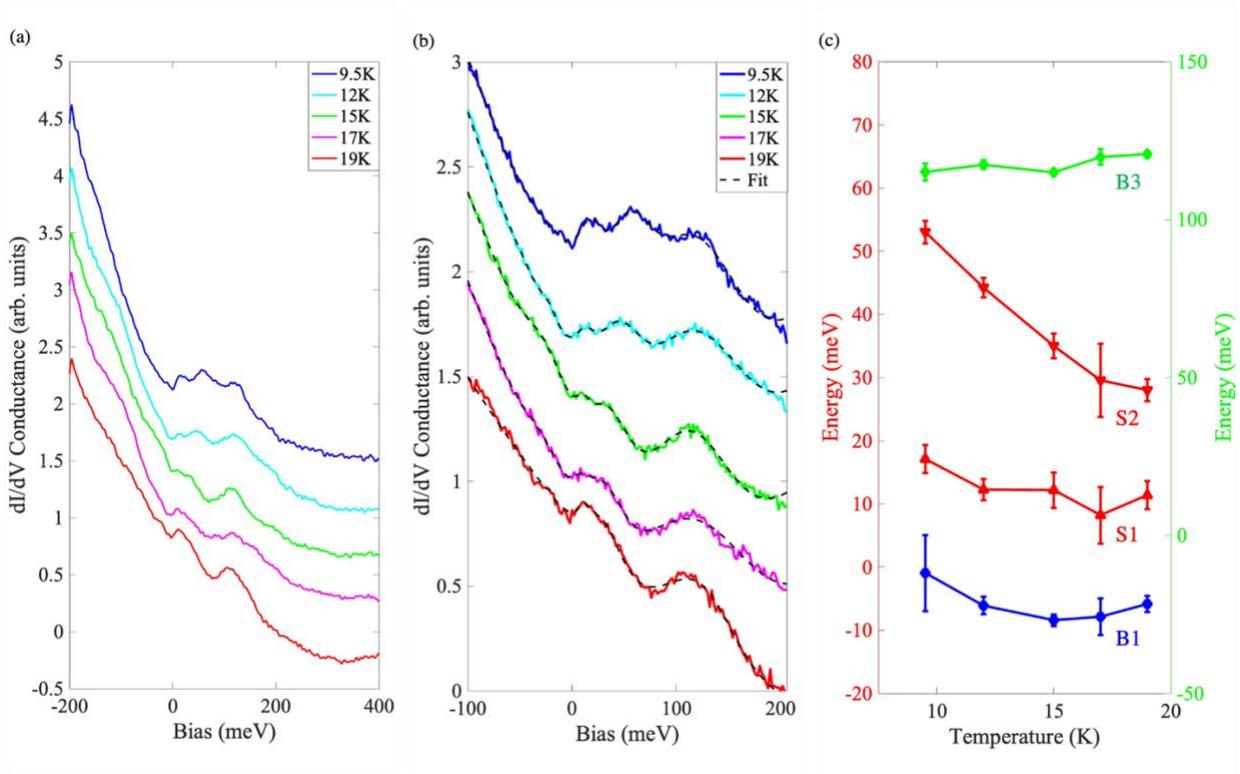

Figure 5. Temperature-dependence of dI/dV conductance spectra on the atomic surface with energy ranges of (a) -200 to 400 meV and (b) -100 to 200 meV. All spectra are taken at the exact same location on the atomic surface with a setpoint bias of -200 meV and a setpoint current of 100 pA. Two dashed lines in (b) track the peak positions of the peaks corresponding to the surface state bands, showing the trend of shrinking magnetic gap as temperature increases with the peaks eventually merged together at 19K, above $T_N$. The black dashed lines correspond to a Gaussian fitting with four Gaussian functions for the B1, S1, S2, and B3 peaks and a quadratic background. The red and green vertical dashed lines guide the eye to the evolution of the peaks. Note that the B1 hump is quite weak since it is composed predominantly of In character (see main text). (c) Energy versus temperature of the four Gaussian peak energies extracted from the fits in (b). The SS red triangles correspond to the left y-axis while the bulk bands B1 and B3 correspond to the right y-axis. The vertical arrow corresponds to the magnitude of the full bandwidths of the S1 and S2 peaks extracted from the fits.

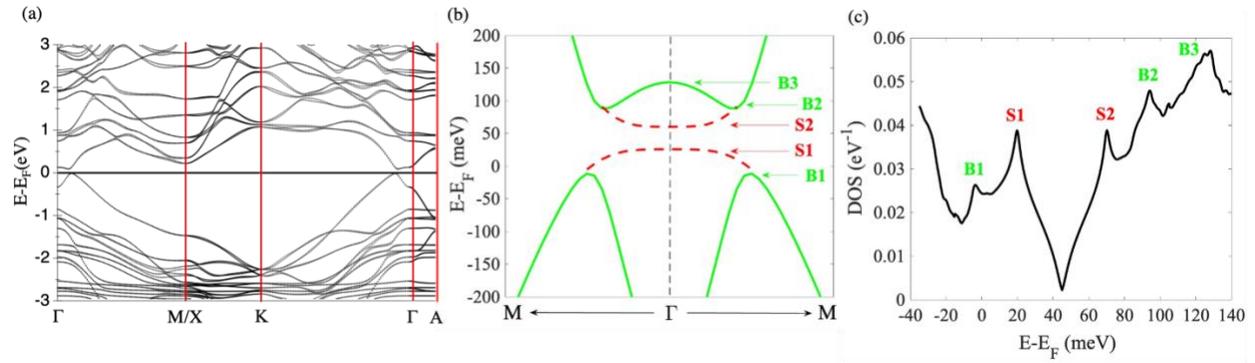

Figure 6: (a) DFT calculated band structure. (b) DFT calculated band structure along the M-Γ-M direction near $E_F$. B1, B2/B3 correspond to Van Hove singularities of the bulk valence and conduction bands respectively, created as a result of spin-orbit induced band-inversion. S1 and S2 correspond to Van Hove singularities of the gapped surface states [44]. (b) The calculated density of states shows peaks (corresponding to Van Hove singularities in (b)) at energies that are in good agreement with the STM spectra observed on the atomic surface (shown in Fig. 3(a)).

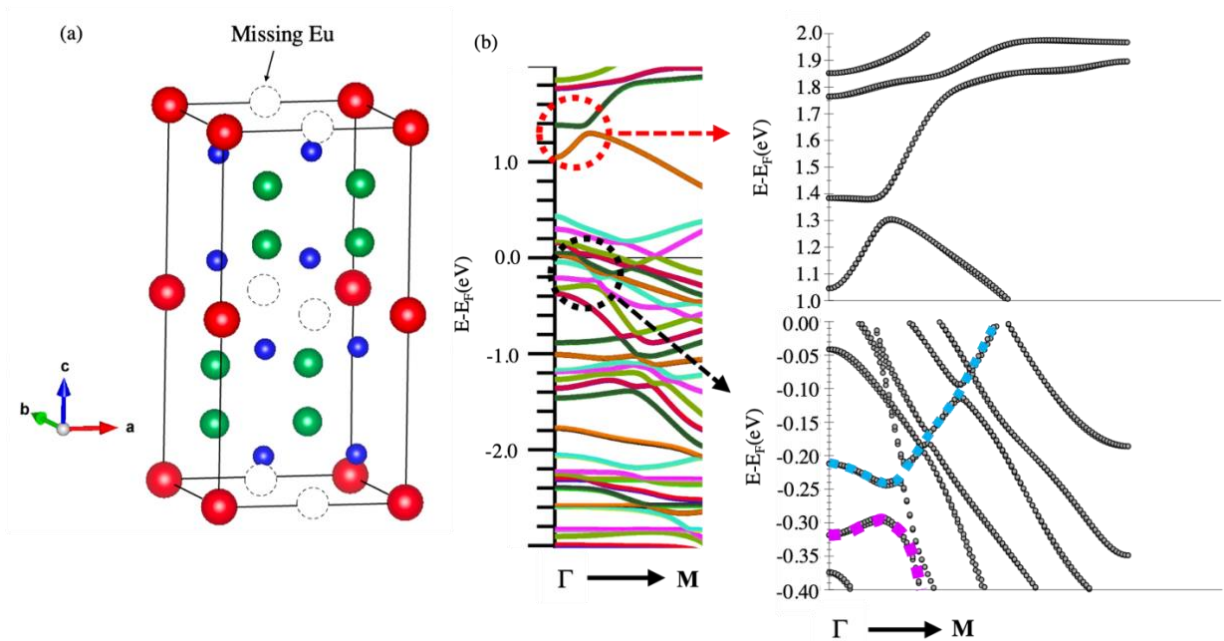

Figure 7. (a) Crystal structure of EuIn2As2 with every other Eu atom missing along the a-direction, resulting in the stripe surface. The sites missing the Eu atom are present in each Eu layer, thus the unit cell is doubled along the a-direction. (b) The band structure around the Gamma point near the Fermi energy. It can be seen that one copy of the band touching point is pushed up to an energy of ~1.3 eV, while the other copy is pushed down to an energy of ~0.25 eV and merged into the Eu f-bands. The splitting between these two copies is way larger because the sites missing Eu atom are present in each Eu layer. This structure represents the case of large splitting due to the stripe order of Eu atoms.

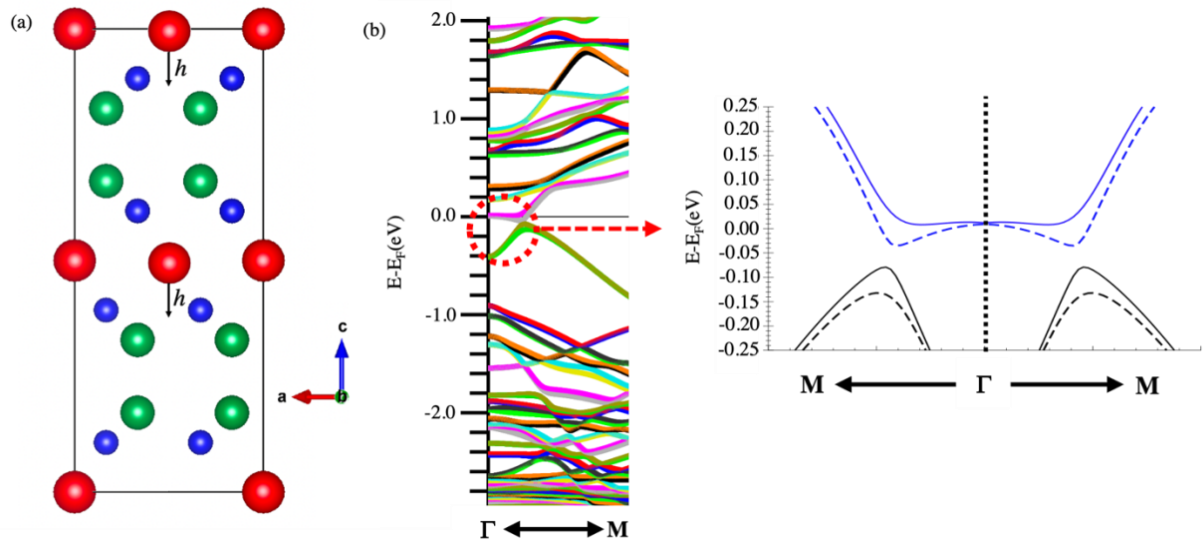

Figure 8: (a) Artificial crystal structure in which one of the Eu atoms is placed closer to its nearby As atom by h=0.2Å. The unit cell is consequently doubled along the a-direction. (b) The DFT band structure around the Gamma point near the Fermi energy of the artificial crystal structure. The band replication can be clearly seen on the right. In this structure, because Eu atoms are only shifted but not missing from the site, the electrical potential inducing the band replication is much smaller, leading to the smaller band splitting. This structure represents the case of small splitting due to the stripe order of Eu atoms.